\documentclass[nameyear]{elsart}
\usepackage{epsfig}

\newcommand{\be}{\begin{eqnarray}}
\newcommand{\ee}{\end{eqnarray}}
\def\lsim{\mathrel{\rlap{\lower3pt\hbox{\hskip1pt$\sim$}}
    \raise1pt\hbox{$<$}}} 
\def\gsim{\mathrel{\rlap{\lower3pt\hbox{\hskip1pt$\sim$}}
    \raise1pt\hbox{$>$}}} 
\newcommand{\msun}{\mbox{$\,M_\odot$}}
\newcommand{\rsun}{\mbox{$\,R_\odot$}}

\def\Ms{}

  \def\astrobjCygX1{Cyg X-1}
  \def\astrobjNovaSco{Nova Scorpii 1994}
  \def\astrobjV444Cyg{V444 Cygni}
  \def\astrobjSN1987A{SN 1987A}
  \def\astrobjLMCXa{LMC X-1}
  \def\astrobjLMCXb{LMC X-3}
  \def\astrobjWR22{WR22}

\begin{document}

\runauthor{Brown, Heger, Langer, Lee, Wellstein, \& Bethe} 

\begin{frontmatter}
\title{Formation of High Mass X-ray Black Hole Binaries}
\author[suny]{G.E. Brown\thanksref{geb}}
\author[ucsc]{A. Heger\thanksref{alex}}
\author[uu]{N. Langer\thanksref{langer}}
\author[suny,kias]{C.-H. Lee\thanksref{chl}}
\author[potsdam]{S. Wellstein\thanksref{wellstein}}
\author[cornell]{H.A. Bethe}

\address[suny]{Department of Physics \& Astronomy,
        State University of New York,\\
        Stony Brook, New York 11794, USA}
\address[ucsc]{Department of Astronomy and Astrophysics,
          University of California,\\ Santa Cruz, CA 95064, USA}
\address[uu]{Astronomical Institute, P.O. Box 80000, NL-3508
          TA Utrecht, The Netherlands}
\address[kias]{School of Physics,
       Korea Institute for Advanced Study, Seoul 130-012, Korea}
\address[potsdam]{Institut f\"ur Physik, Universit\"at Potsdam,
          Am Neuen Palais 10, \\ D-14415 Potsdam, Germany}
\address[cornell]{Floyd R. Newman Laboratory of Nuclear Studies,
          Cornell University,\\ Ithaca, New York 14853, USA}

\thanks[geb]{E-mail: popenoe@nuclear.physics.sunysb.edu}
\thanks[alex]{E-mail: alex@ucolick.org}
\thanks[langer]{E-mail: N.Langer@astro.uu.nl}
\thanks[chl]{E-mail: chlee@kias.re.kr}
\thanks[wellstein]{E-mail: stephan@astro.physik.uni-potsdam.de}


\begin{abstract}
The discrepancy in the past years of many more black-hole soft
X-ray transients (SXTs), of which a dozen have now been identified,
had challenged accepted wisdom in black hole evolution. Reconstruction
in the literature of high-mass X-ray binaries has required stars of up
to $\sim 40\msun$ to evolve into low-mass compact objects, setting this
mass as the limit often used for black hole formation in population
syntheses. On the other hand, the sheer number of inferred SXTs requires 
that many, if not most, stars of ZAMS masses $20-35\msun$ end up as
black holes (Portegies Zwart et al. 1997; Ergma \& van den Heuvel 1998).

In this paper we show that this can be understood by challenging
the accepted wisdom that the result of helium core burning in a massive
star is independent of whether the core is covered by a hydrogen envelope,
or ``naked" while it burns. The latter case occurs in binaries when the
envelope of the more massive star is transferred to the companion by
Roche Lobe overflow while in either main sequence or red giant stage.
For solar metallicity,
whereas the helium cores which burn while naked essentially never go
into high-mass black holes, those that burn while clothed do so,
beginning at ZAMS mass $\sim 20\msun$, the precise mass depending
on the $^{12}$C$(\alpha,\gamma$)$^{16}$O rate as we outline. In this
way the SXTs can be evolved, provided that the H envelope of the massive
star is removed only following the He core burning.

Whereas this scenario was already outlined in 1998 by Brown, Lee, \& Bethe
(1999) their work was based on evolutionary calculations of Woosley,
Langer, \& Weaver (1995) which employed wind loss rates which were too high.
In this article we collect results for lower, more correct wind loss rates,
finding that these change the results only little.

We go into the details of carbon burning in order to reconstruct why
the low Fe core masses from naked He stars are relatively insensitive 
to wind loss rate. 
The main reason is that without the helium produced
by burning the hydrogen envelope, which is convected to the carbon in
a clothed star, a central $^{12}$C abundance of $\sim 1/3$ remains unburned
in a naked star following He core burning. 
The later convective burning
through $^{12}$C$+^{12}$C reactions occurs at a temperature $T\sim 80$ keV.

Finally, we show that in order to evolve a black hole of mass $\gsim 10\msun$
such as observed in \astrobjCygX1, 
even employing extremely massive progenitors of
ZAMS mass $\gsim 60\msun$ for the black hole, the core must be covered by
hydrogen during a substantial fraction of the core burning. In other words,
the progenitor must be a WNL star. We evolve \astrobjCygX1 in an analogous way
to which the SXTs are evolved, the difference being that the companion
in \astrobjCygX1 is more massive than those in the SXTs, 
so that \astrobjCygX1 shines continuously.

\end{abstract}


\end{frontmatter}

\newpage
\section{Introduction}
\label{sec1}

Accepted wisdom in the astrophysical community (for recent relevant
articles see Portegies Zwart, Verbunt, \& Ergma 1997; Ergma \& van den
Heuvel 1998) has been that the results of burning the helium core of a 
star are independent of whether it is covered by a hydrogen envelope or
whether it burns as a ``naked" star, the envelope having been removed;
e.g. by Roche Lobe overflow in main sequence or red giant stage of the
star. This latter situation takes place in binary evolution where the
evolution referred to in the above two articles determines the
black hole progenitors to have ZAMS mass $>40\msun$. Thus, this mass
limit for black hole formation is adopted for the evolution of the
black hole soft X-ray transients (SXTs) in these articles. On the other
hand, as acknowledged in them, the observed number of SXTs requires
that many, if not most, stars of ZAMS masses in the interval of
$\sim 20-35\msun$ evolve into high-mass black holes, of mass $\gsim 6\msun$.

Brown, Lee, \& Bethe (1999) offered as explanation of the large number 
of SXTs that stars in the mass range $\sim 20-35\msun$ could evolve into
high-mass black holes as long as their H envelope was lifted off by the
companion star only after the He core burning was complete (Case C
mass transfer). If it were lifted off earlier, too small an Fe core 
evolved to make a high-mass black hole. In fact, Timmes, Woosley, \& 
Weaver (1996) had found that Fe cores of Type Ib supernovae were
systematically of lower mass than those of Type II, basically the
same concept. The Brown, Lee, \& Bethe scenario had been outlined
by Brown, Weingartner, \& Wijers (1996). Their results were not generally
believed, however, because the calculation of Woosley, Langer, \& Weaver
(1995) on which they relied employed mass loss rates that were too
large by a factor of 2 to 3 for the naked He stars. Thus, the present
work using lower, more correct, rates is necessary.

Wellstein \& Langer (2000) recalculated the evolution of naked He cores
to their carbon-oxygen cores with wind loss rates lowered by a factor of 2,
and in some cases by larger factors, finding that because of feedbacks,
the total wind loss scaled more slowly than linearly with
the multiplayer on the wind loss rate.
In work that we describe here, Alexander Heger, using the Woosley \&
Weaver computer program Kepler, evolved the CO cores further to
their Fe cores. These are not sufficiently massive to end up as high-mass
black holes; in fact, the situation is nearly the same as found by
Brown, Weingartner \& Wijers (1996) where too large He star mass loss
rates had been used. Our main objective in this article is to explain
why the masses of the final Fe cores depend only insensitively
on the He core mass loss rates. We interpret the Fe core mass as a good
indicator of the final compact core mass because binding energy corrections
and mass fallback very nearly compensate each other (Brown, Weingartner,
\& Wijers 1996).

The plan of this paper is as follows:

In Sec.~\ref{sec2} we discuss the improved measurements of wind losses
as deduced from Wolf-Rayet stars.

In Sec.~\ref{sec3} we show that the large differences between ``naked" and
``clothed" He star burning can be understood in terms of the different
ways in which carbon burns in the two cases.

In Sec.~\ref{sec4} we deal with a possible evolution of the $\sim 10\msun$
black hole in \astrobjCygX1, suggesting that there may be an analogy with the
evolution of transient sources; namely that the He core must be kept
covered by H during most of its burning. This requires a very massive
progenitor, a WNL star.

In Sec.~\ref{sec5} we give simple concluding remarks.

\section{Wind Loss Rates in WNE's (naked He stars)}
\label{sec2}

It is the mass loss rate proposed by Langer (1989a) for hydrogen-free
Wolf-Rayet stars, which replaced the previously used assumption of a
constant Wolf-Rayet mass loss rate in many evolutionary calculations
(Schaller et al. 1992; cf. WLW 1995), 
which leads to the small final masses of
stars above $\sim 35\msun$. This semi-empirical rate has been
criticised as too high, one argument being the existence of massive
black hole binaries.  Indeed, recently studies of Wolf-Rayet stars
show that originally measured wind losses have to be corrected
downwards by a factor of 2...3 to account for their ``clumpiness"
(Hamann \& Koesterke 1998). This is supported by polarisation
measurements of the Thomson scattering, which depend linearly on the
wind density (St.-Louis et al. 1993, Moffat \& Robert 1994), and is in
approximate agreement with the rates that would be deduced from the
observed rate of increase in orbital periods for spherical mass loss
    \be
    \frac{\dot M}{M}=
    \frac{\dot P}{2 P} 
    \label{eq1}
    \ee
in Wolf-Rayet binaries.  In \astrobjV444Cyg $\dot P=0.202\pm 0.018\; 
{\rm s} \;
{\rm yr}^{-1}$ was obtained by Khaliullin et al. (1984) and $M_{WR}
=9.3\pm 0.5 \msun$ by Marchenko et al. (1994), resulting in
    \be
    \dot M_{dyn} \simeq 1\times 10^{-5} \msun \ {\rm yr}^{-1}.
    \label{eq2}
    \ee
This is to be compared with the
    \be
    \dot M=0.75\times 10^{-5}\msun \ {\rm yr}^{-1}
    \label{eq3}
    \ee
obtained by St.-Louis et al. (1993) from the polarisation
measurements.  In later work Moffat \& Robert (1994) arrive at a mean
of $(0.7\pm 0.1)\times 10^{-5}\msun \ {\rm yr}^{-1}$.  For WNE
(defined here as a Wolf-Rayet phase with vanishing surface hydrogen
abundance) stars,
WLW (1993) used a He wind rate of
    \be
    \dot M=5\times 10^{-8} (M_{WR}/\msun)^{2.6} \dot{\msun} \ {\rm yr}^{-1}
    \ee
which would give a mass loss rate of $1.6\times 10^{-5} \msun$
yr$^{-1}$ for a $9.3\msun$ WR, a factor of 1.6 to 2.1 greater than
eqs.~(\ref{eq2}) and (\ref{eq3}). Given the errors in measurement, we
feel safe in saying that the true wind loss rate is not less than 1/3
to 1/2 the WLW value. Furthermore, we should mention that absolute
value slope of the
mass loss relation Eq.~(4) agrees well with the latest empirical 
mass loss rate determinations of Nugis \& Lamers (2000). 

However, decreasing the rate to 1/3 does not change our conclusions,
as shown by recent calculations with such a reduced mass loss rates by
Wellstein \& Langer (1999), by Fryer et al. (2001),
and by the calculations of Heger described later in this paper.
These show that stars with initial masses below $60\msun$ are unlikely
to produce black holes more massive than $10\msun$, from naked
helium stars, although high mass black holes ($\sim 5 -10\msun$) may
result from the ZAMS mass range $\sim 20-35\msun$ employed by Brown,
Lee, \& Bethe (1999) provided Case C mass transfer occurs in the
binary. The relative insensitivity of the final Fe core masses to the
wind-loss rate depends somewhat intricately on the way carbon burns
in a naked He core, as we describe in the next section.

Before we go on to the next section we discuss, however, the possible
evolution of \astrobjCygX1, a continuously shining black hole binary where
the black hole mass is observed to be $\gsim 10\msun$. 
Even though massive stars are so prone to lose their hydrogen
envelopes and then suffer extreme wind mass loss as hydrogen-free
Wolf-Rayet stars, Langer (1987) pointed out that the most massive
stars may in fact avoid this kind of evolution. The reason for doing
so is that these stars, upon core hydrogen exhaustion,
produce an extended region of intermediate hydrogen abundance in
between the helium core and the hydrogen-rich envelope. While the
latter is supposed to be lost quickly in a so called Luminous Blue
Variable (LBV) stage (e.g., Stothers 2000), this intermediate region may be
sufficiently massive that the star does not manage to blow it away
during the ensuing so called WNL stage (defined here as a Wolf-Rayet
phase with non-vanishing surface hydrogen abundance).

An example for the resulting evolutionary sequence O~star
$\rightarrow$ LBV $\rightarrow$ WNL $\rightarrow$ SN is the $100\msun$
sequence of Langer \& El~Eid (1986). Also WLW
(1993) showed that the final masses of stars above $\sim 80\msun$
may increase again for larger initial masses, while stars in the range
$\sim 35...60\msun$ obey the opposite trend.  Even using the high mass
loss rate, their $85\msun$ WRB star evolved a final He core of
$9.71\msun$.  We will estimate in Section~\ref{sec4} that this can be
brought up to $\sim 16\msun$ with the lower, more correct, mass loss
rates. The reason that the He core of the WLW $85\msun$ star was so
much more massive than that of their $60\msun$ star, was not only the
higher mass of the progenitor but the fact that the WNL stage of the
$85\msun$ comprised $\sim 40\%$ of the Wolf-Rayet stage, whereas that of the
$60\msun$ star was $\sim 25\%$. During the WNL stage the star is
covered by hydrogen, so that the mass loss rate is smaller than it
would have been for a naked He star, and moreover the helium core mass
increases with time.

Although the evolution of star as massive as $\sim 85\msun$ is very
uncertain, we believe that the only possibility for them to end up in
$\gsim 10\msun$ black holes is an extended WNL stage, lasting a major
fraction of the core helium burning. As we discuss below, these stars
would have an active hydrogen burning shell source, whereas the
existence of such a stage in a single massive star would probably not
affect its evolution substantially, an expansion of the star due to
hydrogen shell burning does give the possibility of common envelope
evolution with a companion.
In other words, the most massive stars are covered by H during much of 
their core He burning time, and this may enable them to end up as
high-mass black holes.

\section{``Naked'' vs ``Clothed'' Helium Core Burning}
\label{sec3}

\subsection{Central Carbon Abundances vs Convective Carbon Burning}

As noted earlier, Timmes et al. (1996) and Brown, Weingartner, \&
Wijers (1996) showed that compact cores which followed from ``naked"
helium stars were substantially less massive than followed from those
``clothed" during helium burning with hydrogen envelopes.  More
recently, this has been explored in detail by Wellstein \& Langer
(1999) and Fryer et al. (2001).  An important property of
the stellar core that is significantly influenced by this difference
is the carbon abundance in the core at central helium depletion.
If its abundance is high enough, central carbon burning become
exothermic enough to significantly overcome the neutrino losses and
burn in a convective core.  Otherwise the first convective burning of
carbon can only appear in a shell source.  The carbon abundance also
has significant effect on the location, duration, and extent of the
carbon burning shells.  For example, as the carbon abundance goes to
zero, no convective shell burning occurs either.  The lower the carbon
abundance, the further out the first shell forms.  The location of the
last carbon-burning shell sets the size of the carbon-free core which
determines the final evolution of the star and the sizes of the iron
core (possible direct black hole formation) and the silicon core
(possible black hole formation by fall-back).


There are two major factors that contribute to the resulting carbon
abundance after core helium depletion.  The first is the dependence on
the mass of the helium core and results chiefly from the different
behavior of the $^{12}$C$(\alpha,\gamma)^{16}$O reaction, as described
clearly by Weaver \& Woosley (1993).  Since this is important to our
development, we repeat the argument, supplemented by more recent
developments.

Carbon is formed in the triple $\alpha$-process; since this is a
three-body process, it depends on the density as $\rho^2$.  Carbon is
removed, when possible, by the $^{12}$C$(\alpha,\gamma)^{16}$O
process, which goes linearly with $\rho$.  
Now, for massive helium cores the central density scales roughly as
$M^{-1/6}$.  With increasing helium core mass this implies that
carbon formation, which goes as $\rho^2$, will be cut down, compared
with $^{12}$C$(\alpha,\gamma)^{16}$O, which goes as $\rho$.  Here, we
also note that both the $^{12}$C$(\alpha,\gamma)^{16}$O reaction and
the triple $\alpha$-process have very similar temperature dependence
in the temperature regime of central helium burning, $2\times10^8\,$K,
i.e., the carbon production does not depend on temperature in this
regime.

This means that there is a mass above which the post-helium burning
central carbon abundance is low enough to skip central carbon burning.
Woosley \& Weaver (1995) found this transition at a mass of
19$\msun$.  However, this limit strongly depends on the still
uncertain $^{12}$C$(\alpha,\gamma)^{16}$O reaction rate and the
stellar evolution model, in particular the prescription used for
convection or mixing processes in general, and also mass loss.  
On this last issue we elaborate later in more detail.

A bare helium star behaves differently from a clothed one in that its
helium core does not grow in mass due to hydrogen shell burning, as is the
case in clothed stars, but rather shrinks due to mass loss from the
surface -- the whole star is only a helium core, a Wolf-Rayet star.
Generally, the convective core tends to comprise a larger fraction of
the helium core, by mass, as the central helium abundance decreases,
due to the density-dependence outlined above.  This statement depends,
again, somewhat on the description of semiconvection, but probably
applies to most models with either Schwarzschild convection (e.g.,
Schaller et al. 1992), overshooting or fast semiconvection (e.g.,
Woosley \& Weaver 1995), or rotationally induced mixing (Heger,
Langer, Woosley 2000).  Additionally, the mass fraction of the
convective core also increases with the mass of helium core.  However,
bare helium cores (early-type Wolf-Rayet stars) experience mass loss
rates that are sufficiently strong that the convective core actually tends
to shrink in the run of its evolution, in particular also towards the
end of helium burning, rather than grow, as it does in a clothed star.
This is despite the fact that
it still comprises an increasingly larger fraction of
the remaining total helium core.  The important point is that
a growth in mass of the convective core injects new helium into this
convective core, while when the mass is constant or decreases, this
injection does not occur.

As the triple alpha reaction depends on the third power of the helium
mass fraction it loses against the $^{12}$C$(\alpha,\gamma)^{16}$O
reaction toward the end of central helium burning, i.e., carbon is
mostly burned rather then produced toward the end of central helium
burning.  That switch typically appears at a central helium
mass fraction of around 10--20\,\%. 
Most importantly, as can be seen from the central carbon abundances
at the end of He burning in Table~\ref{tab1}, the He fraction is too low
to burn the final carbon, which remains at a central abundance $\gsim
0.30$ for all mass stars. This is roughly double the carbon abundance
necessary for convective carbon burning.
It is seen in Table~\ref{tab2} that the central
carbon abundance in naked He stars goes down only slowly with
decreasing mass loss rate, although by the time the mass loss is
reduced by a factor of 6 the convective burning stops.
 In the clothed stars the growth
of the core and its accompanied injection of helium after this time
thus leads to a further decrease of carbon as compared to the bare
helium cores that do not have this additional supply of helium.

In Fig.~\ref{fig1} we show calculations carried out by Tom Weaver
(priv.\ com.\ 1995) using the KEPLER code (Weaver, Zimmerman, \&
Woosley 1978; Woosley \& Weaver 1995) on the central carbon abundance
after core helium depletion for ``clothed'' (single) stars without
(any) mass loss as function of ZAMS mass.  The rapid drop in $C_c$ at
ZAMS mass $M_{ZAMS}\sim 20\msun$ causes the disappearance of
convective carbon burning.

\subsection{Iron Core Masses}

It is generally known that the central entropy rises with the mass of
the star; so that the more entropy is carried off, the less massive
will be the final Fe core. This statement can be made more quantitatively
by using the Bethe, Brown, Applegate, \& Lattimer (1979) conclusion
that the entropy per nucleon in the Fe core is $\lsim 1$, in units
of the Boltzmann constant. For a ZAMS $20\msun$, the central entropy
in our Fe core is 0.76 per nucleon when the He core is burned clothed,
0.775 when burned naked; i.e., essentially the same. Thus, the
total entropy of the Fe core is $\sim N_N$ considering some gradient
towards higher entropy at the outer layers of the core,
where $N_N$ is the number
of nucleons in it. In this way one can see directly that the entropy
that is carried off during the evolution of the Fe core will
diminish its mass.

In Fig.~\ref{fig2} we show iron core masses at the time of iron core
collapse for a finely spaced grid of stellar masses (Heger, Woosley,
Martinez-Pinedo, \& Langanke 2001).  Filled circles and crosses
correspond to the core masses of ``clothed'' single stars.  The
circles were calculated by Tom Weaver (1995, priv.\ com.) using the
same physics as in Woosley \& Weaver (1995), whereas the crosses
employ the improved weak rates by Langanke \& Martinez-P\'{\i}nedo (2000) for
electron capture and beta decay.  Although the latter are much smaller
than the rates by Fuller, Fowler, \& Newman (1985) used by Woosley \&
Weaver (1995), the final Fe core masses are not much changed, for
reasons described by Heger, Woosley, Martinez-Pinedo, \& Langanke
(2001).

For ``clothed'' single stars, one sees an anticorrelation between
the peak in Fe core masses at ZAMS mass $\sim 23\msun$ and the minimum
in $C_c$ at ZAMS mass $\sim 21\msun$. In other words, the peak around
$23\msun$ in ZAMS masses occurs just where the central carbon
abundance at the end of helium core burning is at its minimum of $\sim
10\%$, too low for convective core burning or the formation of a
carbon-burning convective shell close to the center of the star.

When sufficient carbon remains, it will be burned convectively in
reactions such as $^{12}$C$+^{12}$C$\rightarrow ^{24}$Mg, $^{20}$Ne$+
\alpha$ etc., at a temperature of $70-80$ keV, several times higher
than the $\sim 20$ keV needed for $^{12}$C$(\alpha,\gamma)^{16}$O.  If
the carbon abundance is lower, the carbon burning phases are reduced
and typically bigger carbon-free cores result for otherwise similar
stars.  Therefore less energy is carried away by neutrinos and the
entropy in the core stays higher. More massive silicon and iron cores
can form in this case (Boyes, Heger, and Woosley 2001).
This is clearly seen in Figures~\ref{fig1} \& \ref{fig2}. 

Although the
central carbon abundance goes up again for ZAMS mass $\sim 25\msun$, and
the Fe core mass goes down, by such high masses the gravitational
energy of the stellar envelope is large and it is difficult for the
shock energy after collapse to blow it off, so we believe that these
will go into high-mass black holes, as well as the stars of ZAMS mass
$20-23\msun$.

\subsection{Formation of High-Mass Black Holes}

Although there is a lot of uncertainty in the literature about the
ZAMS masses which end up in high mass black holes, Bethe \& Brown
(1999) argued that this should occur at the proto compact core mass of
   \be
   M_{PC}\sim 1.8\msun.
   \ee
Further uncertainty comes in relating the Fe core mass to the compact
core mass. It can, however, be seen from Table 3 of Brown,
Weingartner, \& Wijers (1996) that the estimated fallback mass
following separation in the supernova explosion roughly compensates
for the binding energy increase in going from the Fe core to the
compact object. Thus, we will assume the compact object masses to be
the same as those of the Fe cores.  In addition to the uncertainty in
this relation, there is the further uncertainty that the fallback mass,
which is dependent on the density structure in the layers above the
Fe core (Fryer 1999; Janka 2001), is unknown, since consistent supernova 
explosions have not been achieved.

None the less, we believe it useful to move ahead with our estimated
$M_{PC}$ which indicates that single stars with ZAMS masses $\gsim
20\msun$ go into high-mass black holes. Brown, Lee, \&
Tauris (2001) recently showed that given presently accepted wind
losses (Schaller et al. 1992) the high mass black holes in the
transient sources with main sequence companion can only be evolved in
the region of ZAMS mass $\sim 20\msun$, so our choice of $M_{PC}$ is
supported to this extent by evolutionary arguments.

We note briefly that \astrobjSN1987A, if it formed a black hole, went into a
low mass black hole of mass $1.5\msun <M_{BH} <1.8\msun$ according to
the Brown \& Bethe (1994) estimates. The black hole would have a low
mass in this interval because the He envelope was blown off in the
delayed explosion.  Note from Fig.~\ref{fig2} of main sequence masses
that there is only a narrow interval from $\sim 18\msun$ to $20\msun$
in which this could happen given the above interval.

Schaller et al. (1992) use a $^{12}$C$(\alpha,\gamma)^{16}$O 
S-factor of S(300 keV)
$\sim 100$ keV barns as compared with the Woosley \& Weaver 170 keV
barns.  They bring their central abundance down to 0.16 at the end of
core He burning only for a ZAMS $25\msun$ star, so presumably this
would be the mass at which the Fe cores would begin to rise rapidly in
mass with further evolution. We believe the Weaver \& Woosley value
for the $^{12}$C$(\alpha,\gamma)^{16}$O rate to be more correct
(Boyes, Heger, \& Woosley 2001),
placing the narrow interval of ZAMS masses from which \astrobjSN1987A can be
evolved correctly.
In fact, most recently experiments including both E1 and E2 components
have been carried out (Kunz et al. 2001), obtaining
$S^{300}_{tot}=(165\pm 50)$ keV barns.

\section{Possible Evolution of \astrobjCygX1}
\label{sec4}

\subsection{Space Velocity vs the Mass Loss in the Formation of \astrobjCygX1}

According to the calculations of Wellstein \& Langer (1999) and Fryer
et al. (2001), it is unlikely that a black hole as massive
as the $10\msun$ core (Herrero et al. 1995) in \astrobjCygX1 can be evolved
from a naked He star (see Table~\ref{tab1} and Fig.~\ref{fig2}).  
Although the calculations of naked He stars
were extended up to only ZAMS $60\msun$ stars, the He winds scale with
the 2.6 power of the mass, so that higher mass He stars would be
expected to behave in the same way.  In fact, Wellstein \& Langer
(1999) and Fryer et al. (2001) were unable to evolve a
$10\msun$ black hole, whereas the high space velocity of \astrobjCygX1,
$49\pm 14$ km s$^{-1}$ (Kaper et al. 1999) indicates substantial mass
loss in a Blaauw-Boersma kick (Blaauw 1961; Boersma 1961) in the black
hole formation, increasing the necessary He core mass.  We now
estimate this mass loss, following the development of Nelemans et
al. (1999).
It is reasonable to evolve \astrobjCygX1 analogously to the soft X-ray
black hole transient sources, the difference being in the high mass
companion star which presumably makes \astrobjCygX1 shine continuously.

The high space velocity of \astrobjCygX1 can be explained by mass loss in
the black hole formation, because this loss from the black hole is
somewhat off from the center of gravity of the system.  From Nelemans
et al. (1999), the runaway velocities from symmetric SNe (See Appendix
for the derivation)
   \be
   v_{sys}&=&213 \times \Delta M\times  m\times
   P_{re-circ}^{-1/3}\times  (M_{BH}+m)^{-5/3}
   \ {\rm km\ s^{-1}}
   \ee
where masses are in $\msun$, $P$ in days, and $\Delta M=M_{He}-M_{BH}$
with $M_{He}$ the He star mass of the Black hole progenitor.
Here $P_{re-circ}$ is the re-circularized orbital period
after Blaauw-Boersma kick, and
we assume no orbital evolution ($P_{re-circ}=P_{obs}$)
since the beginning of the mass transfer phase,
and neglect small eccentricity as in Nelemans et al. (1999).
By putting in average values,
   \be
   v_{sys} &=& 8.32 \times \Delta M\ {\rm km\ s^{-1}}.
   \ee
In order to obtain the observed velocity, we need $\Delta M\sim
5.9\msun$, indicating the black hole progenitor mass to be $M_{He}
\sim 16 \msun$. Note that this $\Delta M$ and $M_{He}$ are in the same
ballpark as those estimated by Nelemans et al. for \astrobjNovaSco 
(GRO J1655$-$40). This suggests that the evolution of \astrobjCygX1 is
similar to that of \astrobjNovaSco, except that the progenitor mass of
the black hole must be much higher in \astrobjCygX1 and the companion is an
O-star rather than an F-star.

The current orbital separation of \astrobjCygX1 is
  \be
  a_{\rm now}
  \approx 4.2 \rsun \left(\frac{P_{orb}}{\rm days}\right)^{2/3}
              \left(\frac{M_{BH}+M_O}{\msun}\right)^{1/3}
  \approx 40\rsun .
  \ee
Before the explosion, the orbital separation of the black hole
progenitor and companion star was
  \be
  a_{\rm preSN} \approx \frac{a_{\rm now}}{1+e_2} \approx 33 \rsun
  \ee
where $e_2=\Delta M/(M_{\rm BH}+M_O)\approx 0.21$ is the eccentricity
right after the supernova explosion. For the derivation, see Appendix.
This is only slightly larger than the $30\rsun$ estimated by Bethe
\& Brown (1999) for the minimum initial separation of the O-star
progenitors.  If the progenitors were initially closer together they
would merge already at this stage of evolution.

As noted by Bethe \& Brown (1999), the two O-star progenitors must
have initially had a separation of nearly $33\rsun$. Substantial mass
loss, especially by the WR star would have been expected to widen the
orbit substantially
   \be
   \frac{a_f}{a_i}=\frac{M_i}{M_f}\sim 2-3
   \ee
where $M_i$ and $M_f$ are the initial and final system masses. In
order to tighten the orbit to the $33\rsun$ we find before the massive
star goes into a black hole, there must have been some period of
nonconservative mass transfer.

\subsection{Extended WNL Stage of Black Hole Progenitor}

Langer (1989b) finds that towards the end of core hydrogen burning the
radius of a massive star may be much larger than its radius on the
zero age main sequence. For example, a $100\msun$ star increases its
radius from $13\rsun$ at H ignition by a factor $\sim 4$ to $53\rsun$
at core H exhaustion. In the evolutionary phase between central H- and
He-burning massive stars develop an intermediate fully convective zone
just above the H-burning shell which reaches its maximal spatial
extent and $\sim 18\msun$ in mass for the $100\msun$ star (Langer
1987).  There may be further expansion in a hydrogen shell burning
stage unless large mass loss sets in. But the latter may begin only
later in the LBV stage. In any case, there will be a period before or
in the LBV stage where the hydrogen is being propelled outwards, but
does not yet have enough velocity to escape. A companion to the very
massive star will, if in this region, couple hydrodynamically to the
hydrogen in the manner explicitly worked out by Bethe \& Brown (1999),
and furnish energy to it from a drop in its gravitational binding;
i.e., through some mechanism resembling common envelope evolution. We
cannot be more explicit, because of the uncertainty in the post main
sequence evolution of the very massive stars, but some nonconservative
mass transfer seems to be necessary in the evolution of a binary such
as \astrobjCygX1.  We admit to being in somewhat of a predicament, however,
since we do not want a common envelope evolution which removes the
hydrogen before the helium burning is well underway.

It is clear that the black hole in \astrobjCygX1 cannot have evolved from a
naked He star. An important clue to its possible evolution is offered
by the evolution of WLW (1993) of a ZAMS $85\msun$ star, where the WR
stage began with a WNL stage which took $\sim 40\%$ of the total WR
time. (Their $60\msun$ star also had a WNL stage, but it took only
$\sim 25\%$ of the WR time.) During this time the WR is covered by
hydrogen and the attendant mass loss rate is much lower than that for
a naked He star.

Using the times for each WR stage as WLW 1995, we calculate in
Table~\ref{tab3} the masses for ZAMS $60\msun$ and $85\msun$
stars. Of course there is feedback on these masses from the change in
times of each stage with altered mass loss rate. In order to take the
feedback into account, we use the ratios of $M_{CO}$ cores calculated
by Fryer et al. (2001) for a ZAMS $60\msun$ star for
various mass loss rates to the original WLW (1995) one as plotted in
Fig~\ref{fig3}.

We note from Table~\ref{tab3} that a ZAMS $60\msun$ star with WR
stage mass loss rate reduced by 1/3 loses $1.7\msun$ during its WNL
stage, at an average rate of $1.5\times 10^{-5} \msun$ yr$^{-1}$.
This is an order of magnitude less than it would lose as a WNE (bare)
He star, and explain why the final He star mass can be as large as
$13.5\msun$.  The WNL mass loss rate is $\sim 2.25$ times the main
sequence mass rate for a ZAMS $60\msun$ star of Castor, Abbott, \&
Klein (1975).

In Fig.~\ref{fig3} we plot our ratios from Table~\ref{tab3} for He
star masses alongside the Fryer et al. (2001) ratios of
CO core masses. The divergence in our favored range of $1/2-1/3$ is
not large, so we believe our procedure in calculating masses is
justified. We see from Table~\ref{tab3} that we need a mass loss rate
of $\sim 0.4$ times the WLW (1995) one in order to lose $\sim 6\msun$
in the explosion and to be left with an $\sim 10\msun$ black hole.

\subsection{Discussion on \astrobjCygX1 Type Objects}

We want to point out that there is a strong metallicity dependance in
our model. As the winds of WNL stars are likely radiation driven
(Hamann et al. 2000), we can expect them to weaken with smaller metallicity.
Thus, the lower limit of the initial mass range from which on the the
final stellar mass can grow again will be decreasing.  Therefore, we
expect more \astrobjCygX1 type systems at lower metallicity.
We also would like to point out that there is dependance in mixing
processes, in particular rotation, which can increase the lower limit
of the initial mass range.

Using standard population synthesis, Bethe \& Brown (1999) estimated
that there should be $\sim 7$ \astrobjCygX1 like objects in the Galaxy,
where we see only one (leaving out \astrobjLMCXa \  and \astrobjLMCXb). 
Although the $\gamma$-rays from such an object easily penetrate the
Galactic disc, they would not be distinguished from those from
gamma ray bursts (Maarten Schmidt, private communication 2001).
 
As possible WR progenitor of the \astrobjCygX1 black hole,
we suggest ``\astrobjWR22"; the most massive Wolf-Rayet ever weighed (Rauw et
al. 1996). The minimum Wolf-Rayet mass is $72\msun$ and the mass ratio
to O-star is 2.78, giving an O-star mass of $>26\msun$. 
The spectrum of the WN7 exhibits a
considerable amount of hydrogen ($X_H\sim 40\%$, Hamann et al. 1991,
Crowther et al. 1995), strong WR emission lines and absorption lines
that belong to the WR component (Niemela 1973; Moffat \& Seggewiss
1978). Rauw et al.  (1996) identify the companion as an O-star. With
$\sim 80$ day period, the binary is, however, much too wide in order
to narrow sufficiently in common envelope evolution to produce a \astrobjCygX1
type binary, according to the Bethe \& Brown (1999) estimates.

\section{Conclusion}
\label{sec5}

Our conclusion to all of the above is simple: In order to evolve high-mass
black holes at solar metallicity, the He core of the massive star must 
be covered by hydrogen at least most of the time, while it burns.

For SXTs, the stars in the mass range $\sim 20-35\msun$ could evolve
into high-mass black holes with the condition that their H envelope
was lifted off by the companion star only after the He core burning
was complete. For more massive stars like \astrobjCygX1 type objects, 
an extended WNL stage could drive the formation of high-mass black holes.

\section*{Acknowledgments}

We thank Stan Woosley for many helpful discussions and are indebted to
Tom Weaver for supplying us with his grid of stellar evolution models.
We wish to thank Maarten Schmidt for communication about the
indistinguishability of further \astrobjCygX1 like objects from gamma
ray bursts.
This work was partially supported by the U.S. Department of Energy
under Grant No.  DE--FG02--88ER40388, by the Alexander von
Humboldt-Stiftung through Grant FLF-1065004, by 2000-2001 KIAS
Research Fund, and by the Deutsche
Forschungsgemeinschaft through Grants La~587/15 and La~587/16.

\appendix 
\section*{Appendix. Blaauw-Boersma Kick}

\setcounter{footnote}{0}
\renewcommand{\thesection}{A}

In this Appendix, the details of the Blaauw-Boersma kick 
(Blaauw 1961, Boersma 1961) are
summarized. Before the explosion,\footnote{Subscript 1 (2)
indicates the properties before
     (right after) the explosion, and subscript ``now" indicates the
     current properties observed.}
from Fig.~\ref{fig4}~(a), $R_{He}$, $v_{He}$, and $\omega_1$ are
given as
  \be
  R_{He} &=& \left( \frac{m}{M_{He}+m}\right) a_1 \nonumber\\
  v_{He} &=& 
          \frac{1}{a_1} \left(G m R_{He}\right)^{1/2} \nonumber\\
  \omega_1 &=& 
         \left( \frac{m}{M_{He}+m}\right) 
         \left( \frac{Gm}{R_{He}^3}\right)^{1/2}.
  \ee
After the explosion, we assume that $\Delta M\equiv
M_{He}-M_{BH}$ was lost from the BH progenitor without interacting
with the binary system. The momentum lost from the binary system
should be compensated by the space velocity of the new binary
system.
  \be
  \Delta p =\Delta M \cdot v_{He} = (M_{BH}+m)\cdot  v_{sys}
  \ee
which gives the space velocity of the new binary system (c.m.
motion of B)
  \be
  v_{sys}= \left(\frac{\Delta M}{M_{BH}+m}\right) v_{He}.
  \label{eq.blaauw}
  \ee

Right after the explosion, in the new c.m. frame of BH and O-star
binary, by defining the semi-major axis $d_2$ and the eccentricity
$e_2$, we have
  \be
  \omega_2 &=& \left(\frac{m}{M_{BH}+m}\right)
               \left(\frac{Gm}{d_2^3}\right)^{1/2} \nonumber\\
  R_{BH} &=& \left(\frac{m}{M_{BH}+m}\right) a_1 \nonumber\\
  v_{BH} &=& v_{He}+v_{sys} =\left(\frac{M_{He}+m}{M_{BH}+m}\right) v_{He} .
  \ee
Since the $v_{BH}$ is greater than that of the circular motion
with radius $R_{BH}$
  \be
  v_{BH} &=&\left(\frac{M_{He}+m}{M_{BH}+m}\right)^{1/2}
             \ v_{circ} > v_{circ} \nonumber\\
  v_{circ}&=&\left(\frac{Gm^2}{a_1 (M_{BH}+m)}\right)^{1/2},
  \ee
one can conclude that $R_{BH}$ is at the minimum of the orbital
separation, i.e.
  \be
  R_{BH}= d_2 (1-e_2).
  \ee
Note that $\vec v_{BH}$ and $\vec R_{BH}$ are perpendicular right
after the explosion as in Fig.~\ref{fig4}~(c).

The orbital angular momentum of the BH can be expressed using
$a_2$ and $e_2$
  \be
  l_a &=& \left[ G m \mu^2 d_2 (1-e_2^2)\right]^{1/2}
       =  \left[ G m \mu^2 R_{BH} (1+e_2) \right]^{1/2}
  \ee
or
  \be
  l_b &=& M_{BH} R_{BH} v_{BH} =\mu a_1 v_{BH}
      = \mu  \left(Gm R_{BH}\right)^{1/2}
             \left(\frac{M_{He}+m}{M_{BH}+m}\right)^{1/2}
  \ee
where $\mu$ is the reduced mass
  \be
  \mu =\frac{M_{BH}\ m}{M_{BH}+m}.
  \ee
By equating $l_a=l_b$ we have
  \be
  e_2 = \frac{\Delta M}{M_{BH}+m} .
  \ee

We assume that there is no angular momentum lost during the
circularization process by tidal locking after SN explosion. From
angular momentum conservation
   \be
   l_{now} &=& \left(G m \mu^2 d_{now}\right)^{1/2} =
              \left[G m \mu^2 d_2 (1-e_2^2)\right]^{1/2}  .
   \ee
Therefore, from
   \be
   d_{now} &=& \left( \frac{m}{M_{BH}+m} \right) a_{now}  \nonumber\\
           &=& d_2 (1-e_2^2) = R_{BH} (1+e_2) = \left(\frac{m}{M_{BH}+m}\right) a_1 (1+e_2)\ ,
   \ee
we have the relation of the orbital separations
   \be
   a_{now}=a_1 (1+e_2).
   \ee
Using this relation, one have
   \be
   \omega_1 &=& (1+e_2)^2 \ \omega_{now}  \nonumber\\
   \omega_2 &=&  (1-e_2^2)^{3/2}\ \omega_{now}
   \ee
or the period relations
   \be
   P_{now} = (1-e_2^2)^{3/2} \ P_2 = (1+e_2)^2\ P_1.
   \ee

Now Eq.~(\ref{eq.blaauw}) can be expressed in terms of observed
quanties using
   \be
   a_{now} &=& \left(\frac{P_{now}}{2\pi}\right)^{2/3}
                \left[ G (M_{BH}+m) \vphantom{\frac 12}\right]^{1/3} \nonumber\\
   v_{He} &=& \left(\frac{1+e_2}{a_{now}}\right)^{1/2}
              \left( \frac{G m^2}{M_{He}+m}\right)^{1/2}
           = \left[\frac{1}{a_{now}}
             \left( \frac{G m^2}{M_{BH}+m}\right)\right]^{1/2} \nonumber\\
          &=& \frac{(2\pi G)^{1/3} \ m}{(M_{BH}+m)^{2/3}\ P_{now}^{1/3}} .
   \ee
Finally the Blaauw-Boersma kick velocity becomes
   \be
   v_{sys} &=&  \frac{\Delta M}{M_{BH}+m} v_{He}
          = \frac{(2\pi G)^{1/3} \Delta M\ m}{(M_{BH}+m)^{5/3}\ P_{now}^{1/3}}
     \nonumber\\
      &=& 213 \left( \frac{\Delta M}{\msun}\right)  \left(\frac{m}{\msun}\right)
         \left(\frac{P_{now}}{\rm days}\right)^{-1/3}
         \left(\frac{M_{BH}+m}{\msun}\right)^{-5/3}
          \ {\rm km\ s^{-1}} \ .
   \ee


\newpage

\renewcommand{\thetable}{\arabic{table}}

\begin{table}[ht]
\begin{center}
\begin{tabular}{lccccccc}
\hline
Model & $M_{MS} $ & $M_{comp} $ & $C_c$ 
& $M_{He}$ & $M_{CO}$ & $M_{Fe}$ & Mass Transfer \\
\hline
1s   &    60  & 34     &  0.30 &       &  3.37  &  1.35  & B \\
2s   &    60  & 34     &  0.32 &  4.07 &  3.07  &  1.50  & A$+$AB \\
5s   &    40  & 30     &  0.33 &  3.84 &  2.87  &  1.49  & A$+$AB \\
7s   &    30  & 24     &  0.34 &  3.63 &  2.71  &  1.46  & A$+$AB \\
10s  &    25  & 24     &  0.36 &  3.39 &  2.42  &  1.49  & A$+$AB \\
17s  &    20  & 18     &  0.36 &  3.39 &  2.18  &  1.56  & B      \\
\hline
\end{tabular}\\
\begin{minipage}{2.5in}
$C_c$ : Central Carbon abundance\\
Masses : in unit of $\msun$
\end{minipage}
\end{center}
\caption{Results of Wellstein \& Langer (1999) and Fryer 
et al. (2001) 
with reduced WR mass loss rate, $1/2$ of WLW (1993) value,
at the end of central He burning.
The $M_{Fe}$ is the final Fe core mass.
}
\label{tab1}
\end{table}

\begin{table}[ht]
\begin{center}
\begin{tabular}{lccccc}
\hline
Model & Mass Loss Rate$^\star$ & $C_c$ & $\tau_{C}$ & $M_{CO}[\msun]$ & $M_{Fe}[\msun]$  \\
\hline
1s1   &     1                  &  0.35 & 3800 yrs   &  2.35  &  1.32  \\
1s2   &    1/2                 &  0.30 & 1600 yrs   &  3.37  &  1.35  \\
1s3   &    1/3                 &  0.27 &  700 yrs   &  4.76  &  1.61  \\
1s4   &    1/4                 &  0.25 &  500 yrs   &  5.93  &  1.75  \\
1s6   &    1/6                 &  0.22 &  No        &  8.53  &  1.50  \\
\hline
\end{tabular}\\
\begin{minipage}{3.5in}
$^\star$ Ratio of WR mass loss to that of WLW 1993\\
$C_c$ : Central Carbon abundance\\
$\tau_C$ : Convective carbon core burning time
\end{minipage}
\end{center}
\caption{Results of Wellstein \& Langer (1999) and Fryer et al.
(2001) for different mass loss rates of a ZAMS 60$\msun$
star at the end of central He burning.  Models ``1s\#'' correspond to
the binary system (ZAMS $60\msun$ and $34\msun$) with Case B mass
transfer.}
\label{tab2}
\end{table}


\begin{table}[ht]
\begin{center} 
\begin{tabular}{|l|cccccc|c|}
\hline
              & \multicolumn{6}{c|}{WR Stage Mass Loss Rate} & \\ \cline{2-7}
\raisebox{1.5ex}{$60\msun$ WRB}   &   1      & 1/2     &  0.4    & 1/3     & 1/4     & 1/6     & \raisebox{1.5ex}{$\tau$ ($10^5$ yrs)} \\
\hline \hline
pre WR stage  &  26.5\Ms & 26.5\Ms & 26.5\Ms & 26.5\Ms & 26.5\Ms & 26.5\Ms & \\ \cline{1-7}
post WNL      &  22.5\Ms & 24.2\Ms & 24.5\Ms & 24.8\Ms & 25.1\Ms & 25.6\Ms & \raisebox{1.5ex}{1.13 (25\%)} \\ \cline{1-7}
post WNE      &  11.5\Ms & 15.4\Ms & 16.5\Ms & 17.5\Ms & 18.9\Ms & 20.8\Ms & \raisebox{1.5ex}{1.64 (36\%)} \\ \cline{1-7}
post WC/WO    &  6.65\Ms & 9.60\Ms & 10.7\Ms & 11.6\Ms & 13.2\Ms & 15.5\Ms & \raisebox{1.5ex}{1.77 (39\%)} \\ \hline
\hline
after feedback$^\star$
              &  6.65\Ms & 9.53\Ms & 11.9\Ms & 13.5\Ms & 16.8\Ms & 24.1\Ms &
              \\ \hline
\end{tabular}\\

\vskip 5mm

\begin{tabular}{|l|cccccc|c|}
\hline
              & \multicolumn{6}{c|}{WR Stage Mass Loss Rate} & \\ \cline{2-7}
\raisebox{1.5ex}{$85\msun$ WRB}  &   1      & 1/2     & 0.4 & 1/3 & 1/4 & 1/6 & \raisebox{1.5ex}{$\tau$ ($10^5$ yrs)} \\ \hline \hline 
pre WR stage  & 45.3\Ms & 45.3\Ms & 45.3\Ms & 45.3\Ms & 45.3\Ms & 45.3\Ms & \\ \cline{1-7} 
post WNL      & 34.3\Ms & 40.0\Ms & 40.8\Ms & 41.6\Ms & 42.5\Ms & 43.4\Ms & \raisebox{1.5ex}{1.40 (41\%)} \\ \cline{1-7} 
post WNE      & 14.7\Ms & 20.9\Ms & 23.0\Ms & 24.7\Ms & 27.5\Ms & 31.2\Ms & \raisebox{1.5ex}{1.25 (36\%)} \\ \cline{1-7} 
post WC/WO    &  9.7\Ms & 14.3\Ms & 16.1\Ms & 17.6\Ms & 20.2\Ms & 24.0\Ms & \raisebox{1.5ex}{0.80 (23\%)} \\ \hline \hline 
after feedback$^\star$  &  9.7\Ms & 13.9\Ms & 17.3\Ms & 19.6\Ms & 24.5\Ms & 35.2\Ms &                  \\ \hline
\end{tabular}\\
$^\star$ scaled by the numerical results of Fryer et al.
(Fig.~\ref{fig3})
\end{center}
\caption{Mass-loss-rate-dependences of masses (in $\msun$) at the
different stages of WR stars (models 60WRB and 85WRB of WLW93). The WR stage
mass loss rates are in units of standard rate of WLW93. The same WR stage
times of WLW93 are used for different mass loss rates.}
\label{tab3}
\end{table}


\pagebreak
\renewcommand{\thefigure}{\arabic{figure}}

\begin{figure}[ht]
\centerline{\epsfig{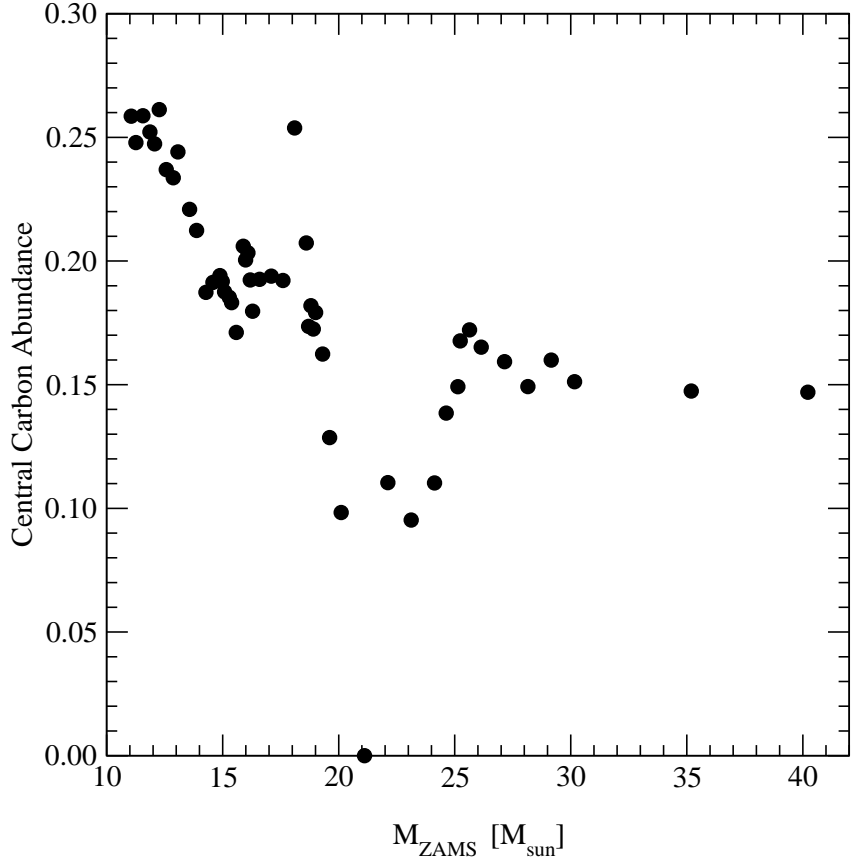}}
\caption{ Central carbon abundance at the end of He core burning for
``clothed" (single) stars as function of ZAMS mass.  The calculation
was carried out by Tom Weaver (1995, priv.\ com.)using the KEPLER code
(Weaver, Zimmerman, \& Woosley 1978; Woosley \& Weaver 1995).  The
rapid drop in $C_c$ at ZAMS mass $M_{ZAMS}\sim 20\msun$ signals the
disappearance of convective carbon burning.  }
\label{fig1}
\end{figure}

\begin{figure}[ht]
\centerline{\epsfig{file=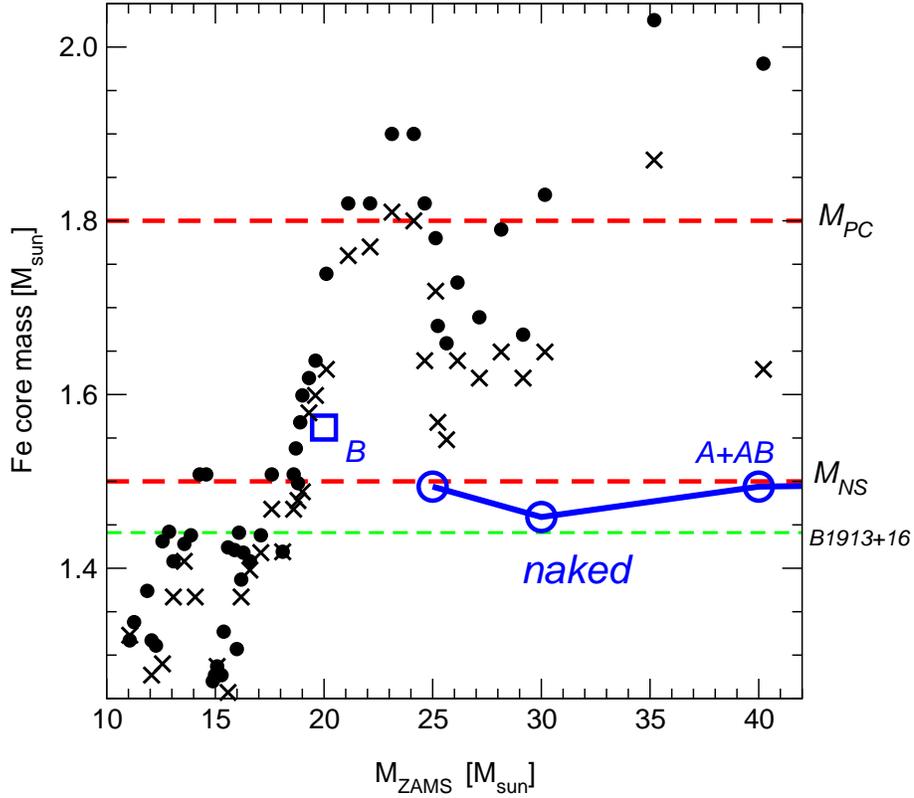,height=5.2in}}
\caption{
Comparison of the iron core masses resulting from
the evolution of ``clothed" and ``naked" He cores. 
Filled circles and crosses correspond to the core masses of ``clothed"
stars at the time of iron core implosion for a finely spaced grid of
stellar masses (Heger, Woosley, Martinez-Pinedo, \& Langanke 2001).
The circular were calculated with the Woosley \& Weaver 1995 code,
whereas the crosses employ the vastly improved Langanke,
Martinez-Pinedo (2000) rates for electron capture and beta decay.
Open circles (square) correspond to the naked He stars in case A$+$AB
(B) mass transfer of Fryer et al. (2001), with reduced WR
mass loss rate, see Table~\ref{tab1}.  If the assembled core mass is
greater than $M_{\rm PC}= 1.8\msun$, where $M_{PC}$ is the
proto-compact star mass as defined by Brown \& Bethe (1994), there is
no stability and no bounce; the core collapses into a high mass black
hole. $M_{NS}=1.5\msun$ denotes the maximum mass of neutron star
(Brown \& Bethe 1994).  The mass of the heaviest known well-measured
pulsar, PSR B1913$+$16, is also indicated with dashed horizontal line
(Thorsett \& Chakrabarty 1999).  }
\label{fig2}
\end{figure}

\begin{figure}[ht]
\begin{center}
\epsfig{file=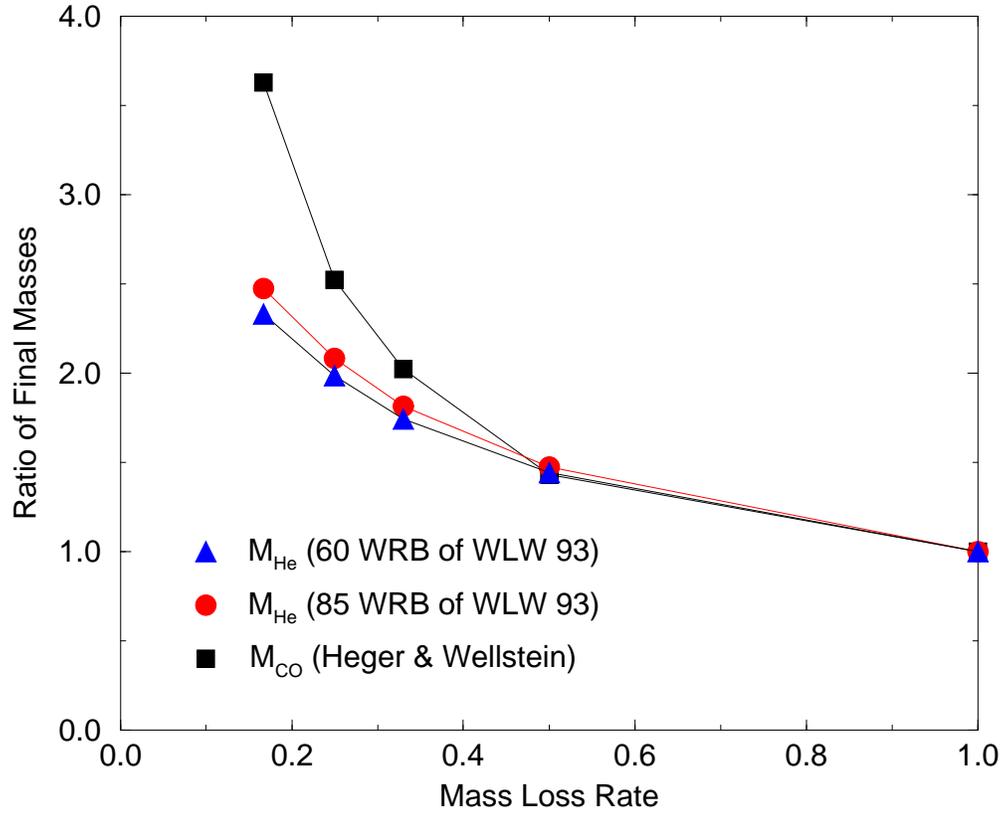,height=5in} \end{center} \caption{
Mass-loss-rate-dependences of final CO core masses M$_{\rm CO}$
(filled square) from the $60\msun$ star discussed in Fryer et al.
(2001) and final He core masses (filled triangles and
circles for $60\msun$ and $85\msun$, respectively) after WR stage from
Tab.~\ref{tab1}. The masses are scaled by those with standard rate of
WLW93.} \label{fig3}
\end{figure}


\begin{figure}[ht]
\centerline{\epsfig{file=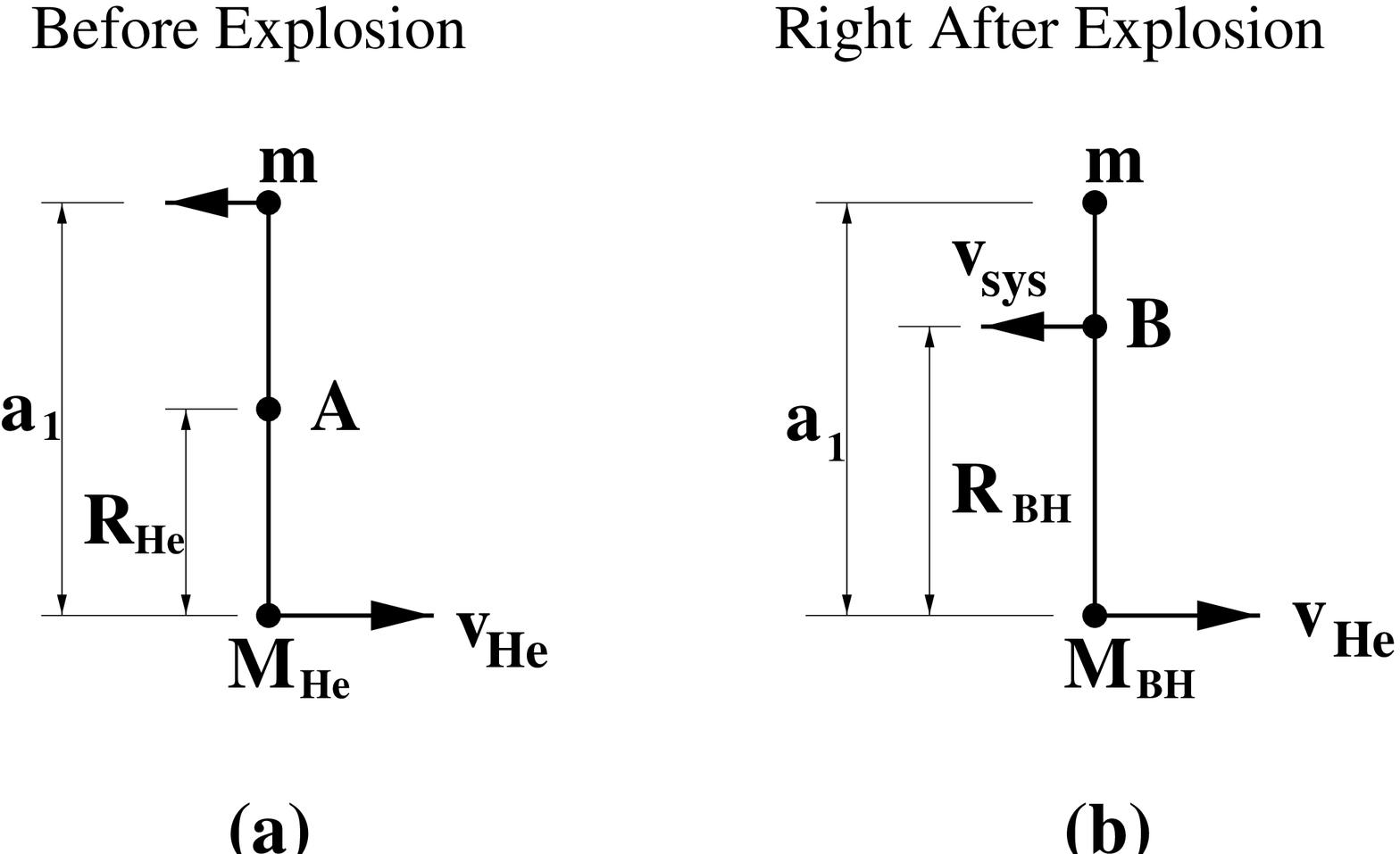,width=3.2in}\ \
\epsfig{file=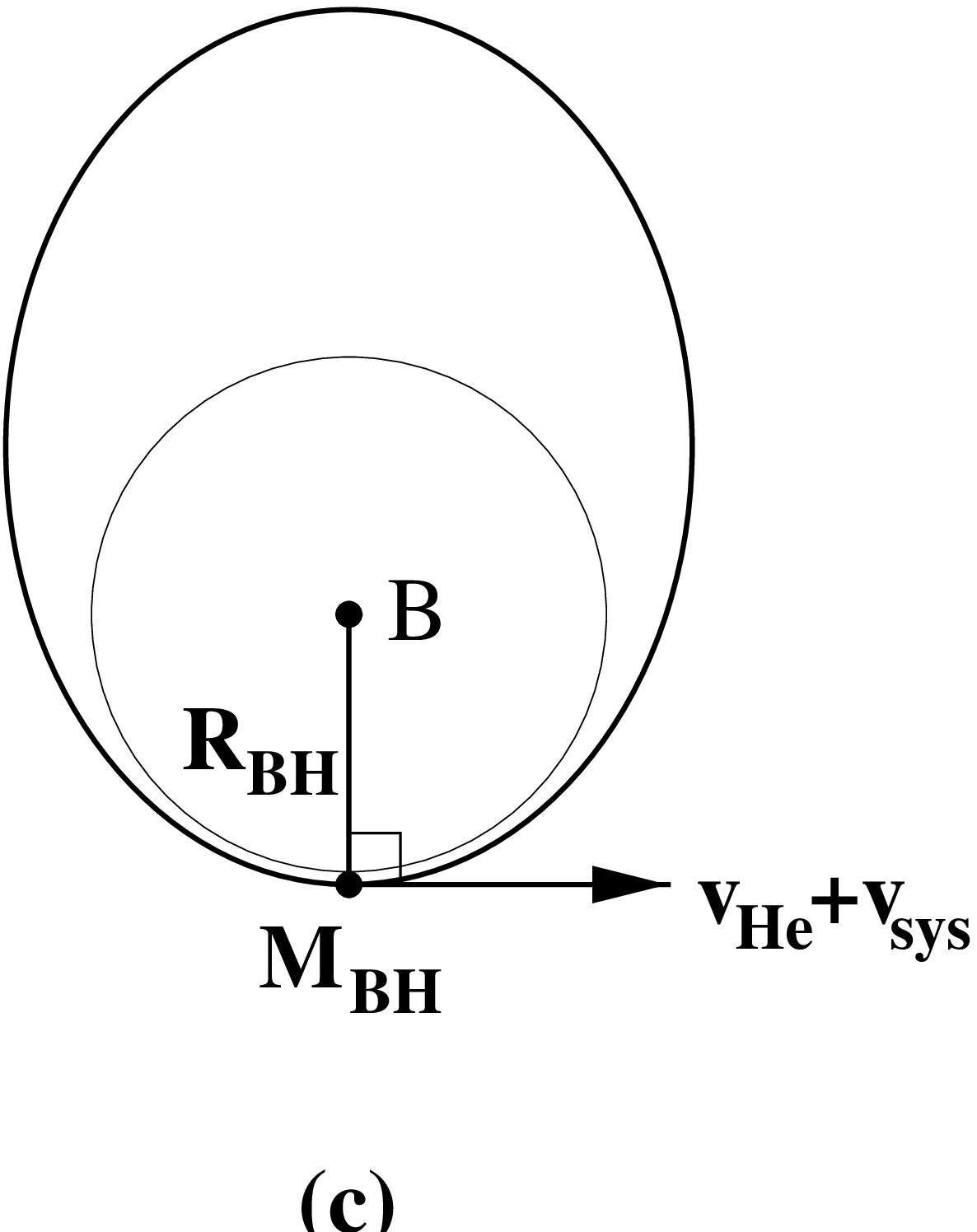,width=1.8in}} 
\caption{Coordinates (a)
before the explosion in the c.m. frame, (b) right after the
explosion in the original c.m. frame, and (c) in the new c.m.
frame of black hole and companion rhght after the explosion. }
\label{fig4}
\end{figure}

\end{document}